\documentclass[11pt]{article} 
\usepackage{amsthm} \usepackage{amsfonts} \usepackage{latexsym}
\newcommand{\braket}[1]{|#1\rangle \langle #1|}
\newcommand{\ket}[1]{|#1 \rangle}
\newcommand{\trace}[1]{\mathrm{Tr}(#1)} \newcommand{\abs}[1]{|#1|}
\newcommand{\norm}[1]{\| #1\|} \newcommand{\alg}[1]{\mathcal{#1}}
\newcommand{\algae}{\mathcal{A}} \newcommand{\alga}{\mathcal{A}}

\newcommand{\weyl}{\mathcal{A}[\mathbb{R}^{2}]} \theoremstyle{plain}
 \newtheorem*{prop*}{Proposition}
 \newtheorem{lemma}{Lemma}
\newtheorem{thm}{Theorem} \newtheorem*{thm*}{Theorem}
 
\theoremstyle{definition} 
 \theoremstyle{remark}

\newcommand{\hil}[1]{\mathcal{#1}}  

\newcommand{\coral}{\mathcal{L}(\psi ,R)}
\newcommand{\bh}{\mathbf{B}(\mathcal{H})} \title{Reconsidering Bohr's
  Reply to EPR\thanks{Forthcoming in J. Butterfield and T. Placek,
    {\it Modality, Probability, and Bell's Theorems}.} }  \author{Hans
  Halvorson\thanks{Philosophy Department, Princeton University; e-mail: hhalvors@princeton.edu} and Rob
  Clifton\thanks{Philosophy Department, University of Pittsburgh;
    e-mail: rclifton@pitt.edu} } \date{October 17, 2001}
\begin{document}
\maketitle
\begin{abstract} 
  Although Bohr's reply to the EPR argument is supposed to be a
  watershed moment in the development of his philosophy of quantum
  theory, it is difficult to find a clear statement of the reply's
  philosophical point.  Moreover, some have claimed that the point is
  simply that Bohr is a radical positivist.  In this paper, we show
  that such claims are unfounded.  In particular, we give a
  mathematically rigorous reconstruction of Bohr's reply to the
  \emph{original} EPR argument that clarifies its logical structure,
  and which shows that it does not rest on questionable philosophical
  assumptions.  Rather, Bohr's reply is dictated by his commitment to
  provide ``classical'' and ``objective'' descriptions of experimental
  phenomena.
\end{abstract}

\section{Introduction}
The past few decades have seen tremendous growth in our understanding
of interpretations of quantum mechanics.  For example, a number of
``no-go'' results have been obtained which show that some or other
interpretation violates constraints that we would expect any plausible
interpretation of quantum mechanics to satisfy.  Thus, although there
is no immediate hope of convergence of opinion on interpretive issues,
we certainly have an increased understanding of the technical and
conceptual issues at stake.  Perhaps, then, we can make use of this
increased technical awareness to shed some new light on the great old
episodes in the conceptual development of quantum mechanics.

One historical episode of enduring philosophical interest is the
debate between Bohr and Einstein (along with Podolsky and Rosen) over
the completeness of quantum mechanics.  Although folklore has it that
Bohr was the victor in this debate, Fine and Beller \cite{fine} have
recently claimed that Bohr's reply to the EPR argument of 1935 is
basically a failure.  In particular, Fine and Beller claim that
``\ldots as a result of EPR, Bohr eventually turned from his original
concept of disturbance, to make a final --- and somewhat forced --- landing in
positivism''~\cite[p.~29]{fine}.  They also make the stronger
philosophical claim that ``\ldots a positivistic shift is the only
salvageable version of Bohr's reply''~\cite[p.~9]{fine}.
Unfortunately, Fine and Beller did not see the need to provide an
argument for this claim.  And, although we are willing to concede ---
for purposes of argument --- that the later Bohr embraced positivism,
we are \emph{not} willing to concede that he was rationally compelled
to do so.  In fact, we will argue that Bohr's defense of the
completeness of quantum mechanics does not depend in any way on
suspect philosophical doctrines.  To this end, we will supply a formal
reconstruction of Bohr's reply to EPR, showing that his reply is
dictated by the dual requirements that any description of experimental
data must be \emph{classical} and \emph{objective}.

The structure of this paper is as follows.  In Section 2, we provide
an informal preliminary account of the EPR argument and of Bohr's
reply.  In Section 3, we consider some salient features of Bohr's
general outlook on quantum theory.  We then return to Bohr's reply to
EPR in Sections 4 and 5.  In Section 4, we reconstruct Bohr's reply to
EPR in the case of Bohm's simplified spin version of the EPR
experiment.  Finally, in Section 5, we reconstruct Bohr's reply to EPR
in the case of the original (position-momentum) version of the EPR
experiment.

\section{Informal Preview}
In classical mechanics, a state description for a point particle
includes a precise specification of both its position and its
momentum.  In contrast, a quantum-mechanical state description
supplies only a statistical distribution over various position and
momentum values.  It would be quite natural, then, to regard the
quantum-mechanical description as \emph{incomplete} --- i.e.~as
providing less than the full amount of information about the particle.
Bohr, however, insists that the imprecision in the quantum-mechanical
state description reflects a fundamental indeterminacy in nature
rather than the incompleteness of the theory.  The EPR argument
attempts to directly rebut this completeness claim by showing that
quantum mechanics (in conjunction with plausible extra-theoretical
constraints) entails that particles always have both a precise
position and a precise momentum.

EPR ask us to consider a system consisting of a pair of spacelike
separated particles.  They then note that, according to quantum
mechanics, there is a state $\psi _{\mathrm{epr}}$ in which the
positions of the two particles are strictly correlated, \emph{and} the
momenta of the two particles are strictly correlated.  It follows then
that if we were to measure the position of the first particle, we
could predict with certainty the outcome of a position measurement on
the second particle; \emph{and} if we were to measure the momentum of
the first particle, we could predict with certainty the outcome of a
momentum measurement on the second particle.

EPR then claim that our ability to predict with certainty the outcomes
of these measurements on the second particle shows that each such
measurement reveals a pre-existing ``element of reality.''  In what
has come to be know as the ``EPR reality criterion,'' they say:
\begin{quote} 
  If, without in any way disturbing a system, we can predict with
  certainty (i.e.~with probability equal to unity) the value of a
  physical quantity, then there exists an element of physical reality
  corresponding to this physical quantity. \cite[p.~77]{epr}
\end{quote}
In particular, if we determine the position of the first particle in
this strictly correlated state, then we can conclude that the second
particle also has a definite position.  And if we determine the
momentum of the first particle in this strictly correlated state, then
the second particle must also have a definite momentum.

Of course, it does not immediately follow that there is any single
situation in which both the position and the momentum of the second
particle are elements of reality.  However, EPR also make the
(\emph{prima facie} plausible) assumption that what counts as an
element of reality for the second particle should be independent of
which measurement is performed on the first particle.  In other words,
a measurement on the first particle can play a \emph{probative}, but
not a \emph{constitutive}, role with respect to the elements of
reality for the second particle.  Consequently, EPR conclude that both
the position and the momentum of the second particle are elements of
reality, regardless of which measurement is performed on the first
particle.

\subsection{Bohr's reply}
According to Bohr, the EPR argument somehow misses the point about the
nature of quantum-mechanical description.  Unfortunately, though, not
much scholarly work has been done attempting to reconstruct Bohr's
reply in a cogent fashion.

We should begin by noting that Bohr most certainly does not maintain
the ``hyperpositivist'' position according to which no possessed
properties or reality should be attributed to an unmeasured system.
(For example, Ruark claims that, for Bohr, ``a given system has
reality only when it is actually measured''~\cite[466]{ruark}.)  Quite
to the contrary, Bohr explicitly claims that when the position of the
first particle is measured, ``\ldots we obtain a basis for conclusions
about the initial position of the other particle relative to the rest
of the apparatus''~\cite[p.~148]{bohr35}.  Thus, Bohr agrees with EPR
that once the position (respectively, momentum) of the first particle
is actually measured, the position of the second particle is an
element of reality --- \emph{whether or not} its position is actually
empirically determined.  In other words, Bohr accepts the outcome of
an application of the EPR reality criterion, so long as its
application is restricted to individual measurement contexts (i.e.~the
results of its application in different contexts are not combined).

In order, then, to rationally reject EPR's conclusion, Bohr must
reject the claim that elements of reality for the second particle
cannot be constituted by measurements carried out on the first
particle.  In other words, Bohr believes that a measurement on the
first particle \emph{can} serve to constitute elements of reality for
the second, spacelike separated, particle.

To this point, we have not said anything particularly novel about
Bohr's reply to EPR.  It is relatively well-known that his reply
amounts to claiming --- what EPR thought was absurd \cite[p.~480]{epr}
--- that what is real with respect to the second particle can depend
in a nontrivial way on which measurement is performed on the first
particle.  However, where previous defenders of Bohr have uniformly
stumbled is in giving a coherent account of \emph{how} a measurement
on one system can influence what is real for some spacelike separated
system.

Unfortunately, Bohr's statements on this issue are brief and obscure.
For example, he says,
\begin{quote}
  It is true that in the measurements under consideration any direct
  mechanical interaction of the [second] system and the measuring
  agencies is excluded, but a closer examination reveals that the
  procedure of measurement has an essential influence on the
  conditions on which the very definition of the physical quantities
  in question rests.  \cite[p.~65]{bohrletter}
\end{quote}
That is, a measurement on the first system influences the conditions
which must obtain in order for us to ``define'' elements of reality
for the second system.  Moreover, this influence is of such a sort
that a position (momentum) measurement on the first particle supplies
the conditions needed to define the position (momentum) of the second
particle.

Before we proceed to our positive account, we need first to dismiss
one \emph{prima facie} plausible, but nonetheless mistaken,
explication of Bohr's notion of defining a quantity.  In particular,
some have claimed that, according to Bohr, an observable of a system
comes to have a definite value when the wavefunction of the system
collapses onto one of that observable's eigenstates.  This amounts to
attributing to Bohr the claim that:
\begin{description} \item[{\it Eigenstate-Eigenvalue Link:}] A quantity $Q$ is defined in state $\psi$
  iff.~$\psi$ is an eigenvector for~$Q$;
\end{description} 
along with the claim that by measuring an observable, we can cause the
quantum state to collapse onto an eigenstate of that observable.  In
that case, Bohr would claim that by measuring the position of the
first particle, we collapse the EPR state onto an eigenstate of
position for the second particle --- and thereby ``cause'' the second
particle to have a definite position.  Similarly, if we were to
measure the momentum of the first particle, we would ``cause'' the
second particle to have a definite momentum.  In either case, the
measurement on the first particle would be the cause of the reality
associated with the second particle.

However, there are at least two good reasons for rejecting this reading
of Bohr.  First, Bohr explicitly claims that a measurement of the
first particle cannot bring about a ``mechanical'' change in the
second particle.  In philosophical terms, we might say that Bohr does
not believe that the position measurement on the first particle
\emph{causes} the second particle to have a position, at least not in
the same sense that a brick can \emph{cause} a window to shatter.
Thus, if Bohr does believe in a collapse the wavefunction, it is as
some sort of \emph{non-physical} (perhaps epistemic) process.
However, it is our firm opinion that, unless the quantum state can be
taken to represent our ignorance of the ``true'' hidden state of the
system, there is no coherent non-physical interpretation of collapse.
(We doubt the coherence of recent attempts to maintain both a
subjectivist interpretation of quantum probabilities, and the claim
that ``there are no unknown quantum states'' \cite{fuchs}.)  Thus, if
Bohr endorses collapse, then he is already committed to the
incompleteness of quantum mechanics, and the EPR argument is
superfluous.

The second, and more important, reason for resisting this reading of
Bohr is the complete lack of textual evidence supporting the claim
that Bohr believed in wavefunction collapse (see \cite{how00}).  Thus,
there is no good reason to think that Bohr's reply to the EPR argument
depends in any way on the notion of wavefunction collapse.

\section{Classical Description and Appropriate Mixtures}
In order to do justice to Bohr's reply to EPR, it is essential that we
avoid caricatured views of Bohr's general philosophical outlook, and
of his interpretation of quantum mechanics.  This is particularly
difficult, because there has been a long history of misinterpretation
of Bohr.  For example, in terms of general philosophical themes, one
might find Bohr associated with anti-realism, idealism, and
subjectivism.  Moreover, in terms of the specific features of an
interpretation of quantum mechanics, Bohr is often associated with
wavefunction collapse, creation of properties/attributes upon
measurement, and ``cuts'' between the microscopic and macroscopic
realms.  However, these characterizations of Bohr are pure distortion,
and can find no justification in his published work.  Indeed, Bohr's
philosophical commitments, and the picture of quantum mechanics that
arises from these commitments, are radically different from the
mythical version that we have received from his critics and from his
well-intended (but mistaken) followers.  (Our own understanding of
Bohr has its most immediate precedent in recent work on ``no
collapse'' interpretations of quantum mechanics
\cite{bub0,bub,bc,beables}.  However, this sort of analysis of Bohr's
interpretation was suggested independently, and much earlier, by Don
Howard \cite{how79}.  See also \cite{how94,how00}.)

According to Bohr, the phenomena investigated by quantum theory cannot
be accounted for within the confines of classical physics.
Nonetheless, he claims that ``\ldots however far the phenomena
transcend the scope of classical physical explanation, the account of
all evidence must be expressed in classical terms''
\cite[p.~209]{bohr49}.  That is, classical physics embodies a standard
of intelligibility that should be exemplified by any description of
the empirical evidence.  In particular, although the various sources
of evidence cannot be reconciled into a single classical picture, the
description of any single source of evidence must be classical.

Bohr's statements about the notion of ``classical description'' have
been horribly misunderstood.  For a catalog of these misunderstandings
and for evidence that they are indeed mistaken, we refer the reader to
\cite{how79,how94,how00}.  On the positive side, we will follow Howard
\cite{how94} in the claim that the notion of classical description is
best explicated via the notion of an ``appropriate mixture.''
\begin{quote} 
  \ldots we make the clearest sense out of Bohr's stress on the
  importance of a classical account of experimental arrangements and
  of the results of observation, if we understand a classical
  description to be one in terms of appropriate mixtures.
  \cite[p.~222]{how94}
\end{quote}
As Howard \cite{how79} shows, the notion of an appropriate mixture can
be developed in such a way that Bohr's (sometimes obscure) statements
about the possibilities of classical description become mathematically
clear statements about the possibility of treating the quantum state
as a classical probability measure.  In order to see this, we first
collect some terminology.

Let $\hil{H}$ be a finite-dimensional vector space with inner-product
$\langle \cdot ,\cdot \rangle$, and let $\bh$ denote the family of
linear operators on $\hil{H}$.  We say that a self-adjoint operator
$W$ on $\hil{H}$ is a \emph{density operator} just in case $W$ has
non-negative eigenvalues that sum to $1$.  If $\psi$ is a vector in
$\hil{H}$, we let $\braket{\psi}$ denote the projection onto the ray
in $\hil{H}$ generated by $\psi$.  Thus, if $\mathrm{Tr}$ denotes the
trace on $\bh$, then $\trace{|\psi \rangle \langle \psi |A}=\langle
\psi ,A\psi \rangle$ for any operator $A$ on~$\hil{H}$.  A
\emph{measurement context} can be represented by a pair $(\psi, R)$,
where $\psi$ is a unit vector (representing the quantum state), and
$R$ is a self-adjoint operator (representing the measured observable).

Following Howard \cite{how79}, we say that a ``mixture,'' represented
by a density operator $W$, is \emph{appropriate} for $(\psi ,R)$ just
in case $W=\sum _{i=1}^{n}\lambda _{i}\braket{\phi _{i}}, \;(n\leq
\mathrm{dim}\hil{H})$, where each $\phi _{i}$ is an eigenvector for
$R$, and $\lambda _{i}=|\langle \psi ,\phi _{i}\rangle |^{2}$ for
$i=1,\dots ,n$.  In other words, $W$ is a mixture of eigenstates for
$R$, and it reproduces the probability distribution that $\psi$
assigns to the values of $R$.  Thus, an appropriate mixture for $(\psi
,R)$ can be taken to represent our ignorance of the value of $R$ in
the state $\psi$.

Once again, we emphasize that Bohr never explicitly invokes
wavefunction collapse, nor does he need to.  Indeed, the idea of a
``measurement problem'' was foreign to Bohr, who seems to take it as a
brute empirical fact --- needing no further explanation from within
quantum theory --- that an observable possesses a value when it is
measured.  Of course, we now know that if Bohr rejects collapse, then
he would also have to reject the claim that an observable possesses a
value only if the system is in an eigenstate for that observable
(i.e.,~the eigenstate$\leftarrow$eigenvalue link) \cite{rob}.  But
there is little reason to believe that Bohr would have been tempted to
endorse this suspect claim in the first place.

\subsection{Appropriate Mixtures and Elements of Reality}
An appropriate mixture is supposed to give a description in which the
measured observable is an ``element of reality.''  However, the
connection between an appropriate mixture (i.e.~some density operator)
and elements of reality is not completely clear.  Clearly, the intent
of writing the appropriate mixture as $W=\sum _{i=1}^{n}\lambda
_{i}\braket{\phi _{i}}$ is that each ``proposition'' $\braket{\phi
_{i}}$ has a truth value.  However, if $W$ is degenerate then $W$ has
infinitely many distinct expansions as a linear combination of
orthogonal one-dimensional projections.  Thus, $W$ itself does not
determine the elements of reality; rather, it is some expansion of $W$
into a linear combination of one-dimensional projections that
determines the elements of reality.

In this case, however, we might as well focus on the one-dimensional
projections themselves.  Thus, we will say that the set $S=\{
\braket{\phi _{i}}:i=1,\dots ,n \}$ is an \emph{appropriate event
  space} for the measurement context $(\psi ,R)$ just in case $S$ is
maximal relative to the following three conditions: (1.) Each $\phi
_{i}$ is an eigenvector of $R$; (2.) If $i\neq j$ then $\phi _{i}$ and
$\phi _{j}$ are orthogonal; (3.) Each $\phi _{i}$ is
\emph{non}orthogonal to $\psi$.  Each of these conditions has a
natural interpretation.  The first condition states that each
proposition in $S$ attributes some value to $R$ (viz.~the eigenvalue
$r_{i}$ satisfying $R\phi _{i}=r_{i}\phi _{i}$); the second condition
states that the propositions in $S$ are mutually exclusive; and the
third condition states that each proposition in $S$ is possible
relative to $\psi$.  Note, moreover, that every appropriate event
space $S$ can be obtained by taking the projection operators in some
orthogonal expansion $W=\sum _{i=1}^{n}\lambda _{i}\braket{\phi _{i}}$
of an appropriate mixture for $(\psi ,R)$, and then eliminating those
projections with coefficient $0$.

If we suppose that $R$ possesses a definite value in the context
$(\psi ,R)$ (and that it is \emph{False} that it possesses any other
value) then an appropriate event space $S$ gives a \emph{minimal} list
of truth-valued propositions in the context $(\psi ,R)$.  However,
Bohr himself is \emph{not} an ontological minimalist; rather, he
claims that ``we must strive continually to extend the scope of our
description, but in such a way that our messages do not thereby lose
their objective and unambiguous character'' \cite[p.~10]{petersen}.
Thus, we should look for the \emph{maximal} set of propositions that
can be consistently supposed to have a truth-value in the context
$(\psi ,R)$.
  
It has been pointed out (in relation to the modal interpretation of
quantum mechanics \cite{kd}) that we can consistently assume that all
projections in $S^{\perp}$ are \emph{False}.  Moreover, if we do so,
then our lattice of truth-valued propositions will be maximal; i.e.~we
cannot add further elements of reality without violating the
requirement of classical description.  Thus, given an appropriate
event space $S$, we will take the full family of truth-valued
propositions to be those in the set (cf.~\cite{kd}):
\[ 
\mathbf{Def}(S):=\Bigl\{ P^{2}=P=P^{*}: \forall Q\in S\,\bigl[ Q\leq P
\;\mbox{or}\; QP=0 \bigr] \;\Bigr\} .\] It is straightforward to
verify that $\mathbf{Def}(S)$ is a sublattice of the lattice of all
projection operators on $\hil{H}$.  Moreover, it can be shown that
$\mathbf{Def}(S)$ is maximal in the following sense: If $\alg{L}$ is a
lattice of projections such that $\mathbf{Def}(S)\subset \alg{L}$,
then $\psi$ \emph{cannot} be represented as a classical probability
distribution over all elements in~$\alg{L}$.  (In this section and the
next, we state results without proof.  Each of these results is a
corollary of the results proved in \cite{beables}.)

\section{Bohr's Reply: Spin Case}
We can now make use of the appropriate mixtures account to reconstruct
Bohr's reply to EPR.  For the sake of mathematical simplicity, we
first consider Bohm's spin version of the EPR experiment.  We return
to the original EPR experiment in the final section.

Suppose that we have prepared an ensemble of spin-$1/2$ particles in
the singlet state:
\begin{equation}
\psi \:=\: \frac{1}{\sqrt{2}} \bigl( \, \ket{x\,+}\ket{x\,-}-\ket{x\,-}\ket{x\,+} \,
\bigr) ,
\end{equation}
where $\sigma _{x}\ket{x\,\pm}=\pm \ket{x\, \pm}$.  Then, $\sigma
_{x}\otimes I$ is strictly anticorrelated with $I\otimes \sigma _{x}$,
and $\sigma _{y}\otimes I$ is strictly anticorrelated with $I\otimes
\sigma _{y}$.  Thus, the outcome of a measurement of $\sigma
_{x}\otimes I$ would permit us to predict with certainty the outcome
of a measurement of $I\otimes \sigma _{x}$; and the outcome of a
measurement of $\sigma _{y}\otimes I$ would permit us to predict with
certainty the outcome of a measurement of $I\otimes \sigma _{y}$.

For any orthonormal basis $\{ \phi _{i}\}$ of eigenvectors for $\sigma
_{x}\otimes I$, the event space \begin{equation} \bigl\{
  \,\braket{\phi _{i}}\,: \, |\langle \psi ,\phi _{i}\rangle |^{2}\neq
  0 \bigr\} ,\end{equation} is appropriate for $(\psi ,\sigma
_{x}\otimes I)$.  However, since $\sigma _{x}\otimes I$ is degenerate,
there are infinitely many distinct orthonormal bases of eigenvectors
for $\sigma _{x}\otimes I$.  Moreover, each basis gives rise to a
distinct event space, and each distinct event space permits us to
attribute \emph{different} elements of reality to the second
(unmeasured) particle.

More concretely, let $P^{x}_{\pm}$ denote the projection onto the ray
generated by $\ket{x\,\pm}$, and similarly for $P^{y}_{\pm}$ and
$P^{z}_{\pm}$.  Then, each of the following event spaces is
appropriate for $(\psi ,\sigma _{x}\otimes I)$:
\begin{eqnarray}
& S_{xx} = \{ P^{x}_{+}\otimes P^{x}_{-} \;,\; P^{x}_{-}\otimes P^{x}_{+}
\} . \nonumber \\
& S_{xy} = \{ P^{x}_{+}\otimes P^{y}_{+} \;,\; P^{x}_{+}\otimes
P^{y}_{-} \;,\; P^{x}_{-}\otimes P^{y}_{+} \;,\; P^{x}_{-}\otimes P^{y}_{-} \} . \nonumber \\
& S_{xz} = \{ P^{x}_{+}\otimes P^{z}_{+} \;,\; P^{x}_{+}\otimes P^{z}_{-}
\;,\; P^{x}_{-}\otimes P^{z}_{+} \;,\; P^{x}_{-}\otimes P^{z}_{-} \} . \nonumber  
\end{eqnarray} 
Clearly, though, these event spaces give \emph{theoretically
  inequivalent} descriptions of the measurement context.  While
$S_{xx}$ gives a description in which the second particle has spin-$x$
values that are perfectly anticorrelated with the spin-$x$ values of
the first particle, $S_{xy}$ gives a description in which the second
particle has spin-$y$ values that are uncorrelated with the spin-$x$
values of the first particle.  How do we determine which description
is the \emph{correct} one?

One might be inclined to argue that it is an advantage to have more
than one ``interpretation'' (i.e.~empirically adequate description)
of the same measurement context.  That is, one might argue that there
is no single correct description of the second particle in this
context; rather, there are several incompatible, but individually
acceptable, descriptions of the second particle.  However --- despite
his otherwise unorthodox philosophical stance --- Bohr is not a
pluralist about descriptions of measurement contexts.  Indeed, he
claims that a measurement context uniquely dictates an interpretation.
\begin{quote} 
  \ldots we are not dealing with an incomplete description
  characterized by the arbitrary picking out of different elements of
  reality at the cost of sacrificing other such elements, but with a
  rational discrimination between essentially different experimental
  arrangements and procedures\ldots . \cite[p.~148]{bohr35}
\end{quote}
Thus, the theorist is not free to make a willy-nilly choice of which
elements of reality to ascribe to the second particle; rather, her 
choice is to be fixed (in some, yet to be explicated, way) by the 
measurement context. 

For Bohr, the correct description of the present measurement context
(in which spin-$x$ is measured on the first particle and no
measurement is performed on the second particle) is $S_{xx}$, where
the two particles have perfectly anticorrelated spin-$x$ values.
However, we do not yet have any explanation for \emph{why} Bohr thinks
that this description is privileged.  In the next two sections, we
shall provide an explanation for Bohr's preference.

\subsection{The EPR reality criterion}
Isn't it \emph{obvious} that $S_{xx}$ is the correct description of
the context in which $\sigma _{x}\otimes I$ is measured in the EPR
state?  In particular, if we know that $\sigma _{x}\otimes I$ has some
value (either $+1$ or $-1$) can we not infer immediately that
$I\otimes \sigma _{x}$ has the opposite value?  But what reason do we
have to think that $I\otimes \sigma _{x}$ has any value at all?  Since
we are refusing to invoke wavefunction collapse, it does not help to
note that L{\"u}ders' rule entails that a measurement of $\sigma
_{x}\otimes I$ collapses $\psi$ onto either $\ket{x\,+}\ket{x\,-}$ or
$\ket{x\,-}\ket{x\,+}$.  Perhaps then our intuition that $I\otimes
\sigma _{x}$ has a value is based on some variant of the EPR reality
criterion: If we can predict with certainty the outcome of a
measurement of $I\otimes \sigma _{x}$, then it must possess a value.

According to Howard, it is a contextualized version of the EPR reality
criterion that dictates which properties Bohr attributes to the second
(unmeasured) particle.  Howard says, ``\ldots there is no obvious
reason why, with the added necessary condition of a restriction to
specific experimental contexts, [Bohr] could not accept the EPR
reality criterion as it stands'' \cite[p.~256]{how79}.  He then spells
out concretely what such a contextualized version of the reality
criterion would require.
\begin{quote} 
  Once the experimental context is stipulated, which amounts to the
  specification of the candidates for real status, our decision as to
  which particular properties to consider as real will turn on the
  question of predictability with certainty.  \cite[p.~256]{how79}
\end{quote}
We will now give a formal description of this notion of a
contextualized reality criterion.  

First, when Howard says that the experimental context specifies the
``candidates'' for real status, he presumably means that an observable
must be compatible with the measured observable in order to be such a
candidate.  For example, if we measure $\sigma _{x}\otimes I$, then
$\sigma _{y}\otimes I$ is not even a candidate for real status.
However, in order for the quantum state $\psi$ to be representable
as a classical probability distribution over two projections $P$ and
$P'$, it is \emph{not} necessary for $P$ and $P'$ to be compatible.
Rather, $\psi$ can be represented as a classical probability
distribution over $P$ and $P'$ if and only if $[P,P']\psi =0$.  Thus,
since we wish to maintain that each spectral projection of $R$ has a
truth-value in the context $(\psi ,R)$, we will say that a property
$P$ is a \emph{candidate for real status} just in case $[P,P']\psi =0$
for every spectral projection $P'$ of $R$. 

However, since compatibility (or compatibility relative to a state) is
not transitive, not every observable that is compatible with the
measured observable can be an element of reality.  For example, both
$I\otimes \sigma _{x}$ and $I\otimes \sigma _{y}$ are compatible with
$\sigma _{x}\otimes I$, but it is not possible for both $I\otimes
\sigma _{x}$ and $I\otimes \sigma _{y}$ to be elements of reality.
Thus, we need a criterion that will permit us to choose among the
candidates for real status in such a way that we do not end up with a
set of properties that cannot be described classically.

According to Howard, ``our decision as to which particular properties
to consider as real will turn on the question of predictability with
certainty.''  In other words, $P$ is real only if it is strictly
correlated with one of the possible outcomes of a measurement of $R$;
i.e.~there is some spectral projection $P'$ of $R$ such that $P$ and
$P'$ are strictly correlated in the state $\psi$.  That is, $\langle
\psi ,(P-P')^{2}\psi \rangle =0$, which is equivalent to $P\psi
=P'\psi$.

Let $\alg{R}$ denote the family of spectral projections for $R$.  Then
Howard's proposal amounts to attributing reality to the following set
of properties in the context $(\psi ,R)$: \[ \coral := \Bigl\{
P^{2}=P=P^{*}: [P,R]\psi =0\:\; \& \:\; \exists P' \in \alg{R} \;
\mbox{s.t.} \; P'\psi = P\psi \; \Bigr\} . \] We leave the following
straightforward verifications to the reader: (1.)~$\coral$ is a
sublattice of the lattice of all projections on $\hil{H}$.  (2.)  The
quantum state $\braket{\psi}$ is a mixture of dispersion-free states
on $\coral$.  (For this, recall that it is sufficient to show that
$[P,Q]\psi =0$ for all $P,Q\in \coral$.)  (3.)  $I\otimes
P^{x}_{\pm}\in \alg{L}(\psi ,\sigma _{x}\otimes I)$ and $I\otimes
P^{y}_{\pm}\not\in \alg{L}(\psi ,\sigma _{x}\otimes I)$; and similarly
with the roles of $x$ and $y$ interchanged.

Thus, the contextualized reality criterion accurately reproduces
Bohr's pronouncements on the EPR experiment.  However, there is a
serious difficulty with this analysis of Bohr's reply.  In particular,
the EPR reality criterion is best construed as a version of
``inference to the best explanation'' (cf.~\cite[p.~72]{redhead}): The
best explanation of our ability to predict the outcome of a
measurement with certainty is that the system has some pre-existing
feature that we are detecting.  However, since Bohr is not a classical
scientific realist (see, e.g.,~\cite{how79}), we cannot expect him to be
persuaded by such inferences to the best explanation.  Thus, although
Howard's contextualized reality criterion gives the right answers, it
fails to give a plausible explanation of why Bohr gave the answers he
did.

\subsection{Objectivity and invariance}
Despite Bohr's rejection of classical scientific realism, he maintains
that our descriptions of experimental phenomena must be ``objective.''
Presumably, Bohr's notion of objectivity is to some extent derivative
from the idealist philosophical tradition, and therefore has
philosophical subtleties that go far beyond the scope of this paper.
For our present purposes, however, it will suffice to use a
straightforward and clear notion of objectivity that Bohr might have
endorsed: For a feature of a system to be objective, that feature must
be invariant under the ``relevant'' group of symmetries.  We now
explicate this notion, and we show that it dictates a unique classical
description of the EPR experiment.

Recall that the event space $S_{xy}=\{ P^{x}_{+}\otimes
P^{y}_{+},P^{x}_{+}\otimes P^{y}_{-},P^{x}_{-}\otimes
P^{y}_{+},P^{x}_{-}\otimes P^{y}_{-}\}$ allows us to describe an
ensemble in which the first particle has spin-$x$ values, and second
particle has (uncorrelated) spin-$y$ values.  Now, consider the
symmetry $U$ of the system defined by the following mapping of
orthonormal bases:
\[ \begin{array}{lll} 
\ket{y\,+}\ket{y\, +} \: & \longmapsto & \:+\ket{z \,-}\ket{z\,-} \\
\ket{y\,+}\ket{y\,-} \: & \longmapsto & \:-\ket{z\,-}\ket{z\,+} \\
\ket{y\,+}\ket{y\, -} \: & \longmapsto & \:-\ket{z\, -}\ket{z\,+} \\  
\ket{y\,-}\ket{y\,-} \: & \longmapsto & \:+\ket{z\,+}\ket{z\,+} .\end{array} \]
Then, $U^{*}(\sigma _{x}\otimes I)U=\sigma _{x}\otimes I$, and $U\psi
=\psi$.  That is, $U$ leaves both the state and the measured
observable of the context invariant.  However,  
\[
U^{*}(P_{\pm}^{x}\otimes P_{\pm}^{y})U \:=\: P_{\pm}^{x}\otimes
P_{\pm}^{z} .\] That is, $U$ does not leave the individual elements of
$S_{xy}$, nor even the set as a whole, invariant.  In fact, there is
no quantum state that is dispersion-free on both $P^{x}_{+}\otimes
P^{y}_{+}$ and on its transform $U^{*}(P^{x}_{+}\otimes
P^{y}_{+})U=P^{x}_{+}\otimes P^{z}_{+}$.  Thus, the candidate elements
of reality in $S_{xy}$ are not left invariant by the relevant class of
symmetries.

In general, let us say that a set $S$ of projections on $\hil{H}$ is
\emph{definable} in terms of $\psi$ and $R$ just in case: For any
unitary operator $U$ on $\hil{H}$, if $U\psi =\psi$ and $U^{*}RU=R$
then $U^{*}PU=P$ for all $P\in S$.  It is straightforward to verify
that the set $S_{xx}=\{ P^{x}_{+}\otimes P^{x}_{-} \, ,\,
P^{x}_{-}\otimes P^{x}_{+} \}$ is definable in terms of $\psi$ and
$R$.  In fact, it is the only such appropriate event space for this
context.

\begin{thm} $\{ P^{x}_{+}\otimes P^{x}_{-} \, ,\,
  P^{x}_{-}\otimes P^{x}_{+} \}$ is the unique appropriate event space
  for $(\psi ,\sigma _{x}\otimes I)$ that is definable in terms of
  $\psi$ and $\sigma _{x}\otimes I$.  \end{thm}

\begin{proof} Suppose that $S$ is an appropriate event space for $(\psi
  ,R)$ that is definable in terms of $\psi$ and $R$, and let
  $\braket{\phi}\in S$.  Let $P_{1}=(P^{x}_{+}\otimes
  P^{x}_{-})+(P^{x}_{-}\otimes P^{x}_{+})$ and let
  $P_{2}=(P^{x}_{+}\otimes P^{x}_{+})+(P^{x}_{-}\otimes P^{x}_{-})$.
  Then $U:=P_{1}-P_{2}$ is a unitary operator.  It is obvious that
  $U^{*}(\sigma _{x}\otimes I)U=\sigma _{x}\otimes I$ and $U\psi
  =\psi$.  Thus, definability entails that $\braket{\phi}$ commutes
  with $U$; and therefore $\braket{\phi}$ is either a subprojection of
  $P_{1}$ or is a subprojection of $P_{2}$.  However, the latter is
  not possible since $P_{2}\psi =0$.  Thus, $\braket{\phi}$ is a
  subprojection of $P_{1}$.  However, there are only two
  one-dimensional subprojections of $P_{1}$ that are compatible with
  $R$, namely $P^{x}_{+}\otimes P^{x}_{-}$ and $P^{x}_{-}\otimes
  P^{x}_{+}$.  Since $\braket{\phi}$ must be compatible with $R$ it
  follows that either $\braket{\phi}=P^{x}_{+}\otimes P^{x}_{-}$ or
  $\braket{\phi}=P^{x}_{-}\otimes P^{x}_{+}$.  \end{proof}
  
Thus, we have a situation analogous to simultaneity relative to an
inertial frame in relativity theory.  In that case, there is only one
simultaneity relation that is invariant under all symmetries that
preserve an inertial observer's worldline \cite{malament}.  Thus, we
might wish to regard this simultaneity relation as the correct one
relative to that observer, and the others as spurious.  In the
quantum-mechanical case, there is only one set of properties that is
invariant under all the symmetries that preserve the quantum state and
the measured observable.  So, we should regard these properties as
those that possess values relative to that measurement context.

It is easy to see that $\alg{L}(\psi ,\sigma _{x}\otimes
I)=\mathbf{Def}(S_{xx})$.  Thus Howard's suggestion of applying a
contextualized reality criterion turns out to be (extensionally)
equivalent to requiring that the elements of reality be definable in
terms of $\psi$ and $R$.  It follows that those attracted by Howard's
analysis of Bohr's response to EPR now have independent grounds to think that
$\alg{L}(\psi ,\sigma _{x}\otimes I)$ gives
the correct list of elements of reality in the context $(\psi ,\sigma
_{x}\otimes I)$.

\section{Bohr's Reply: Position-Momentum Case} 
 There are a couple of formal obstacles that we encounter in attempting
to reconstruct Bohr's reply to the \emph{original} EPR argument.
First, there is an obstacle in describing the EPR experiment itself:
The EPR state supposedly assigns dispersion-free values to the
relative position $Q_{1}-Q_{2}$ and to the total momentum
$P_{1}+P_{2}$ of the two particles.  However, $Q_{1}-Q_{2}$ and
$P_{1}+P_{2}$ are continuous spectrum observables, and no standard
quantum state (i.e.,~density operator) assigns a dispersion-free value
to a continuous spectrum observable.  Thus, in terms of the standard
mathematical formalism for quantum mechanics, \emph{the EPR state does
  not exist}.  Second, there is an obstacle in applying the account of
appropriate mixtures to the EPR experiment: Since the position (or
momentum) observable of the first particle has a continuous spectrum,
no density operator $W$ is a convex combination of dispersion-free
states of the measured observable.  Thus, there are no appropriate
mixtures (in our earlier sense) for this measurement context.

We can overcome both of these obstacles by expanding the state space
of our system so that it includes eigenstates for continuous spectrum
observables.  To do this rigorously, we will employ the
$C^{*}$-algebraic formalism of quantum theory.  We first recall the
basic elements of this formalism.

A $C^{*}$-algebra $\alg{A}$ is a complex Banach space with norm
$A\mapsto \norm{A}$, involution $A\mapsto A^{*}$, and a product
$A,B\mapsto AB$ satisfying:
\begin{equation}
(AB)^{*}=B^{*}A^{*} ,\qquad \quad \norm{A^{*}A}=\norm{A}^{2} , \qquad \quad \norm{AB}\leq
\norm{A}\norm{B} 
.\end{equation}
We assume that $\alg{A}$ has a two-sided identity $I$.  Let $\omega$ 
be a linear functional on $\alg{A}$.  We say that
$\omega$ is a \emph{state} just in case $\omega$ is positive
[i.e.~$\omega (A^{*}A)\geq 0$ for all $A\in \alg{A}$], and $\omega$ is
normalized [i.e.~$\omega (I)=1$].  A state $\omega$ is said to be
\emph{pure} just in case: If $\omega =\lambda \rho +(1-\lambda)\tau$
where $\rho ,\tau$ are states of $\alg{A}$ and $\lambda \in (0,1)$,
then $\omega =\rho =\tau$.  A state $\omega$ is said to be
\emph{dispersion-free} on $A\in \alg{A}$ just in case $\omega
(A^{*}A)=\abs{\omega (A)}^{2}$.  If $\omega$ is
dispersion-free on $A$ for all $A\in \alg{A}$, we say that $\omega$ is
\emph{dispersion-free} on the algebra $\alg{A}$.  

We represent a measurement context by a pair $(\omega ,\alg{R})$,
where $\omega$ is a state of $\alg{A}$, and $\alg{R}$ is a mutually
commuting family of operators in $\alg{A}$ (representing the measured
observables).  We are interested now in determining which families of
observables can be described classically as possessing values in the
state $\omega$.  Thus, if $\alg{B}$ is a $C^{*}$-subalgebra of
$\alg{A}$, we say that $\omega |_{\alg{B}}$ (i.e.,~the restriction of
$\omega$ to $\alg{B}$) is a \emph{classical probability measure} (or
more briefly, \emph{classical}) just in case
\begin{equation}
\omega (A)= \int \omega _{\lambda}(A)d\mu (\lambda ) , \qquad A\in
\alg{B} ,\end{equation}
where each $\omega _{\lambda}$ is a dispersion-free state of $\alg{B}$.

We now construct a specific $C^{*}$-algebra that provides the model
for a single particle with one degree of freedom.  In the standard
Hilbert space description of a single particle, we can take our state
space to be the Hilbert space $L_{2}(\mathbb{R})$ of (equivalence
classes of) square-integrable functions from $\mathbb{R}$ into
$\mathbb{C}$.  The position observable can be represented by the
self-adjoint operator $Q$ defined by $Q\psi (x) = x\cdot \psi (x)$ (on
a dense domain in $L_{2}(\mathbb{R})$), and the momentum observable
can be represented by the self-adjoint operator $P=-i(d/dx)$ (also
defined on a dense domain in $L_{2}(\mathbb{R})$).  (We set $\hbar =1$
throughout.)  We may then define one-parameter unitary groups by
setting $U_{a} :=\exp \{ iaQ \}$ for $a\in \mathbb{R}$, and
$V_{b}:=\exp \{ ibP \}$ for $b\in \mathbb{R}$.  Let $\weyl$ denote the
$C^{*}$-subalgebra of operators on $L_{2}(\mathbb{R})$ generated by
$\{ U_{a}:a\in \mathbb{R} \} \cup \{ V_{b}:b\in \mathbb{R} \}$.  We
call $\weyl$ the \emph{Weyl algebra} for one degree of freedom.

Of course, $\alg{A}[\mathbb{R}^{2}]$ itself does not contain either
$Q$ or $P$.  However, the group $\{ U_{a}:a\in \mathbb{R}\}$ can be
thought of as a surrogate for $Q$, in the sense that a state $\omega$
of $\weyl$ should be thought of as an ``eigenstate'' for $Q$ just in
case $\omega$ is dispersion-free on the set $\{ U_{a} :a\in \mathbb{R}\}$.
Similarly, the group $\{ V_{b}:b\in \mathbb{R}\}$ can be thought of as
a surrogate for $P$.  (Moreover, the indeterminacy relation between
$Q$ and $P$ can be formulated rigorously as follows: There is no state
of $\weyl$ that is simultaneously dispersion-free on both $\{ U_{a}:a\in
\mathbb{R}\}$ and $\{ V_{b}:b\in \mathbb{R}\}$
\cite[p.~455]{rindler}.)

\subsection{Formal model of the EPR experiment}
In the standard formalism, the state space of a pair of particles
(each with one degree of freedom) can be taken as the tensor product
Hilbert space $L_{2}(\mathbb{R})\otimes L_{2}(\mathbb{R})$.
Similarly, in the $C^{*}$-algebraic formalism, the algebra of
observables for a pair of particles (each with one degree of freedom)
can be represented as the tensor product $\weyl \otimes \weyl$.

The EPR state is supposed to be that state in which $Q_{1}-Q_{2}$ has
the value $\lambda$, and $P_{1}+P_{2}$ has the value $\mu$.  (At
present, we have no guarantee of either existence or uniqueness.)
Since $\exp \{ ia(Q_{1}-Q_{2})\} =U_{a}\otimes U_{-a}$ and $\exp \{
ib(P_{1}+P_{2})\} =V_{b}\otimes V_{b}$, and since dispersion-free
states preserve functional relations, the EPR state should assign the
(dispersion-free) value $e^{ia\lambda}$ to $U_{a}\otimes U_{-a}$ and 
the value
$e^{ib\mu}$ to $V_{b}\otimes V_{b}$.  Fortunately, for a fixed pair
$(\lambda ,\mu)$ of real numbers, there is a unique pure state
$\omega$ of $\weyl \otimes \weyl$ that satisfies these two conditions
\cite[Theorem 1]{lmp}.  We will simply call $\omega$ \emph{the EPR
  state}.

Suppose then that we are in a context in which all elements in
$\alg{Q}_{1}:=\{ U_{a}\otimes I:a\in \mathbb{R}\}$ can be assigned
definite numerical (complex) values (e.g.,~a context in which the
position of the first particle has been determined).  We can then ask:
Which observables can be consistently described, along with the elements
of $\alg{Q}_{1}$, as possessing values in the state $\psi$?  Since the
elements of $\alg{Q}_{2}:=\{ I\otimes U_{a}:a\in \mathbb{R}\}$ commute
pairwise with the elements of $\alg{Q}_{1}$, we could provide a
consistent description in which the second particle has a definite
position (that is strictly correlated with the first particle's
position).  However, since the elements of $\alg{P}_{2}:=\{ I\otimes
V_{a}:a\in \mathbb{R}\}$ also commute pairwise with the elements of
$\alg{Q}_{1}$, we could provide a consistent description in which the
second particle has a definite momentum (which is uncorrelated with
the position of the first particle).  The requirement of consistency
does not itself tell us which of these descriptions is the correct one.
In order to find a basis for choosing between the descriptions, we
turn again to symmetry considerations.

Let $\alg{A},\alg{B}$ be $C^{*}$-algebras, and let $\pi$ be a mapping
of $\alg{A}$ into $\alg{B}$.  We say that $\pi$ is a
$*$-\emph{homomorphism} just in case $\pi$ is linear, multiplicative,
and preserves adjoints.  If $\pi$ is also a bijection, we say that
$\pi$ is a $*$-\emph{isomorphism}; and we say that $\pi$ is a
$*$-\emph{automorphism} when we wish to indicate that $\alg{B}$ was
already assumed to be isomorphic to $\alg{A}$.  Finally, let $\omega$
be a state of $\alg{A}$, let $\alg{R}$ be a mutually commuting family
of operators in $\alg{A}$, and let $\alg{B}$ be a $C^{*}$-subalgebra
of $\alg{A}$.  We that $\alg{B}$ is \emph{definable} in terms of
$\omega$ and $\alg{R}$ just in case: For any $*$-automorphism $\alpha$
of $\alg{A}$, if $\alpha (\alg{R})=\alg{R}$ and $\omega \circ \alpha
=\omega$, then $\alpha (\alg{B})=\alg{B}$.  Thus, in our present
circumstance, we wish to determine which (if any) of the candidate
algebras of ``elements of reality'' identified above is definable in
terms of the EPR state and the measured observables $\alg{Q}_{1}$.

\subsection{The uniqueness theorem}
We turn now to the main technical result of our paper.  Our main
result shows that there is a unique (subject to the constraint of
maximality) algebra of observables $\alg{B}$ such that: (1.)  It is
consistent to suppose that all elements of $\alg{B}$ possess a
definite value in the EPR state; (2.) The position observable of the
first particle (more precisely: its surrogate unitary group) lies in
$\alg{B}$; and (3.)  $\alg{B}$ is left invariant by all symmetries
that leave the EPR state and the position of the first particle
invariant.  Furthermore, we show that \emph{these requirements alone
  entail that the second particle also has a definite position}.  We
take this result as demonstrating that if the position of the first
particle is assumed to be definite, then the \emph{only} invariant
classical description is one in which the second particle also has a
definite position.

\begin{thm} 
  Let $\omega$ be the EPR state.  There is a unique subalgebra
  $\alg{B}$ of $\weyl \otimes \weyl$ that is maximal with respect to
  the three conditions:
\begin{enumerate}
\item $\omega |_{\alg{B}}$ is a classical probability distribution;
\item $\alg{Q}_{1}\subseteq \alg{B}$;
\item $\alg{B}$ is definable in terms of $\omega$ and $\alg{Q}_{1}$.
\end{enumerate}
Moreover, it follows that:
\begin{enumerate}
\item[4.] $\alg{Q}_{2}\subseteq \alg{B}$.
\end{enumerate}
\end{thm}
\noindent By symmetry, the result also holds if we replace $\alg{Q}_{i}$ with
$\alg{P}_{i}$ throughout the statement of the theorem.

For the proof of the theorem, we will need to invoke two technical
lemmas.  First, let $\bh$ denote the algebra of bounded linear
operators on the Hilbert space $\hil{H}$.  If $\alg{B}$ is a subset of
$\bh$, we let $\alg{B}'$ denote the set of all operators in $\bh$ that
commute with each operator in $\alg{B}$, and we let
$\alg{B}''=(\alg{B}')'$.

\begin{lemma} 
  Let $\alg{B}$ be a $C^{*}$-algebra of operators acting on $\hil{H}$.
  Let $U_{t}=\exp \{ -itH\}$, where $H$ is a bounded self-adjoint
  operator acting on $\hil{H}$.  If $U_{t}\alg{B}U_{-t}=\alg{B}$ for
  all $t\in \mathbb{R}$, then there is a one-parameter unitary group
  $\{ V_{t}:t\in \mathbb{R}\} \subseteq \alg{B}''$ such that
  $U_{t}AU_{-t}=V_{t}AV_{-t}$ for all $A\in \alg{B}$ and $t\in
  \mathbb{R}$.  \label{borchers}
\end{lemma}

\begin{proof} See Theorem 4.1.15 of \cite{sakai}. \end{proof}

\begin{lemma}
  Let $\alga$ be a $C^{*}$-algebra, let $\omega$ be a state of
  $\alga$, and let $(\pi ,\hil{H},\Omega )$ be the GNS representation
  of $\alga$ induced by $\omega$ \cite[p.~279]{kr}.  Suppose that
  $\alpha$ is a $*$-automorphism of $\alga$ such that $\omega \circ
  \alpha = \omega$.  Then there is a unitary operator $U$ on $\hil{H}$
  such that $U\Omega =\Omega$ and $U\pi (A)U^{*}=\pi (\alpha (A))$ for
  all $A\in \alga$.
  \label{spatial}
\end{lemma}

\begin{proof} See Proposition 7.4.12 of \cite{pedersen}. \end{proof}

\begin{proof}{Proof of the Theorem}  Let $\alg{A} =\weyl \otimes
  \weyl$.  By the GNS construction (see \cite[Thm.~4.5.2]{kr}), there
  is a Hilbert space $\hil{H}$, a unit vector $\Omega \in \hil{H}$,
  and a $*$-homomorphism $\pi$ from $\alg{A}$ into $\bh$ such that
  \begin{equation}
\omega (A)=\langle \Omega ,\pi (A)\Omega \rangle ,\qquad A\in
\alg{A}. \end{equation}
Since $\alg{A}$ is simple, $\pi$ is a $*$-isomorphism.  Thus, we
can suppress reference to $\pi$, and suppose that $\alg{A}$ is given
as a $C^{*}$-algebra of operators acting on $\hil{H}$.  We will need
to make frequent use of the following result: For any subalgebra
$\alg{B}$ of $\alg{A}$, $\omega |_{\alg{B}}$ is a classical
probability distribution if and only if $[A,B]\Omega =0$ for all
$A,B\in \alg{B}$ \cite[Prop.~2.2]{beables}.  

Our proof now splits into two parts: (I.) We show that if a subalgebra
of $\alg{A}$ maximally satisfies conditions 1.--3. of the
theorem, then it also satisfies condition 4.  (II.) We show that there
is a unique subalgebra of $\alg{A}$, viz., 
$\alg{F}^{\Omega}\cap \alg{A}$ (to be defined later),
 that maximally satisfies 1.--4.  
To finish off the argument, we note that if $\alg{B}$ is any 
subalgebra of $\alg{A}$ 
satisfying 1.--3., then (by an application of Zorn's lemma) it is
contained in an algebra $\alg{C}$ that maximally satisfies 1.--3., and
hence also maximally satisfies 1.--4., in virtue of part (I.).  
Thus, by part (II.), $\alg{C}=\alg{F}^{\Omega}\cap \alg{A}$, and 
$\alg{B}\subseteq \alg{F}^{\Omega}\cap \alg{A}$, establishing that 
the latter is also the unique subalgebra of $\alg{A}$ maximally 
satisfying just 1.--3. 

(I.) Suppose that $\alg{B}$ is a $C^{*}$-subalgebra of $\alg{A}$ that
maximally satisfies conditions 1.--3. of the theorem.  We wish to show that
$I\otimes U_{a}\in \alg{B}$ for all $a\in \mathbb{R}$.  If we set
  \begin{eqnarray}
    A & := & (1/2)\bigl[ (I\otimes U_{a})+(I\otimes U_{-a})\bigr] , \\
    B & := & (i/2)\bigl[ (I\otimes U_{-a})-(I\otimes U_{a})\bigr] ,
\end{eqnarray} 
then $A+iB=I\otimes U_{a}$.  Thus, it will suffice to show that
$A,B\in \alg{B}$.  We will treat the case of $A$; the case of $B$ can
be dealt with by a similar argument.  Let
\begin{eqnarray}
A' &=& (1/2)[e^{-ia\lambda }(U_{a}\otimes I)+e^{ia\lambda}(U_{-a}\otimes
I)] .\end{eqnarray}
A straightforward calculation (using the definition of the EPR 
state $\omega$ and the fact $\omega (U_{a}\otimes U_{b})=0$ if 
$b\not=-a$ \cite[Eqn.~19]{lmp})
shows that $\omega ((A'-A)^{*}(A'-A))=0$.  Thus, in the GNS representation, 
\begin{eqnarray}
\norm{(A'-A)\Omega }^{2} &=& \langle (A'-A)\Omega ,(A'-A)\Omega
\rangle \\
&=& \langle \Omega ,(A'-A)^{*}(A'-A)\Omega \rangle \:=\:0 .\end{eqnarray} 
Let $H=A'-A$ and let 
$U_{t}=\exp \{ -itH\}$ for all $t\in \mathbb{R}$.
We claim now that if $U_{t}\in
\alg{B}$ for all $t\in \mathbb{R}$ then $A\in \alg{B}$.  Indeed,
suppose that $U_{t}\in \alg{B}$ for all
$t\in \mathbb{R}$.  Since $\lim _{t\rightarrow
  0}\norm{iH-t^{-1}(U_{t}-I)}=0$, 
and since $\alg{A}$ is closed in the norm topology,
it follows that $H\in \alg{B}$.  Moreover, since $A=H-A'$ and $A'\in
\alg{B}$, it follows that $A\in \alg{B}$.  Thus, it will suffice to
show that $U_{t}\in \alg{B}$ for all $t\in \mathbb{R}$.
  
For each $t\in \mathbb{R}$, define a $*$-automorphism $\alpha _{t}$ of
$\alga$ by setting $\alpha _{t}(Z) = U_{t}ZU_{-t}$ for all $Z\in
\alga$.  Since $H\Omega =(A'-A)\Omega =0$, it follows that
$U_{t}\Omega =\exp \{ -itH \}\Omega =\Omega$ for all $t\in
\mathbb{R}$.  Thus,
\begin{equation}
\omega (\alpha _{t}(Z))= \langle \Omega ,U_{t}ZU_{-t}\Omega \rangle =
\langle \Omega ,Z\Omega \rangle =\omega (Z)
,\end{equation}
for all $Z\in \alga$.  Moreover, $\alpha _{t}(X)=U_{t}XU_{-t}=X$ for all $X\in
\alg{R}$.  Since $\alg{B}$ is definable 
in terms of $\alg{R}$ and
$\omega$, it follows that $U_{t}\alg{B}U_{-t}=\alpha _{t}(\alg{B})=
\alg{B}$ for all $t\in
\mathbb{R}$.  Thus, Lemma \ref{borchers} entails that there is a 
unitary group $\{ V_{t}:t\in \mathbb{R}\}\subseteq
\alg{B}''$ such that $U_{t}ZU_{-t}=V_{t}ZV_{-t}$
for all $Z\in \alg{B}$ and $t\in \mathbb{R}$.  Since $\omega |_{\alg{B}''}$ is
classical (see \cite[Cor.~2.9]{beables}), and since $Z ,V_{-t}\in \alg{B}''$, we have
\begin{eqnarray}
U_{t}ZU_{-t}\Omega=V_{t}ZV_{-t}\Omega=V_{t}V_{-t}Z\Omega=Z\Omega
.\end{eqnarray}
Thus, $[U_{t},Z]\Omega=0$ for all $t\in \mathbb{R}$.    

Let $[\alg{B}\Omega]$ denote the closed linear span of $\alg{B}\Omega
=\{ Y\Omega :Y\in \alg{B} \}$, and let $P$ denote the orthogonal
projection onto $[\alg{B}\Omega]$.  Then $P\in \alg{B}'=(\alg{B}'')'$,
and $\alg{B}''P$ is a von Neumann algebra acting on $[\alg{B}\Omega]$
\cite[Prop.~5.5.6]{kr}.  Let $\alg{B}^{\Omega}$ denote the subalgebra
of $\bh$ given by
  \begin{equation} \alg{B}^{\Omega}=(I-P)\bh (I-P)\oplus \alg{B}''P
    .\end{equation}  
In order to complete the first part of the proof, we show 
(a.) $\alg{B}=\alg{B}^{\Omega}\cap \alga$, and (b.) $U_{t}\in \alg{B}^{\Omega}\cap
\alg{A}$ for all $t\in \mathbb{R}$.  
  
(a.) Recall that $\alg{B}$ was assumed to be maximal with respect to
conditions 1.--3. of the theorem.  Since $\alg{B}\subseteq
\alg{B}^{\Omega}\cap \alg{A}$, it will follow that
$\alg{B}=\alg{B}^{\Omega}\cap \alg{A}$ if it can be shown that
$\alg{B}^{\Omega}\cap \alg{A}$ satisfies conditions 1. and~3.

We first show that $\omega |_{\alg{B}^{\Omega}\cap \alg{A}}$ is a
classical probability distribution.  Since $\omega |_{\alg{B}}$ is
classical, and $\omega$ is a normal state in the representation, it
follows that $\omega |_{\alg{B}''}$ is classical.  Thus, $\alg{B}''P$
is abelian \cite[Prop.~2.2]{beables}, and $\omega
|_{\alg{B}^{\Omega}}$ is classical \cite[Thm.~2.8]{beables}.
Therefore $\omega |_{\alg{B}^{\Omega} \cap \alg{A}}$ is classical.

We now show that $\alg{B}^{\Omega} \cap \alg{A}$ is definable in terms
of $\alg{R}$ and $\omega$.  Let $\alpha$ be a $*$-automorphism of
$\alga$ such that $\alpha (\alg{R})=\alg{R}$ and $\omega \circ \alpha
=\omega$.  Since $\alg{B}$ is definable in terms of $\alg{R}$ and
$\omega$, $\alpha (\alg{B})=\alg{B}$.  By Lemma \ref{spatial}, there
is a unitary operator $U$ on $\hil{H}$ such that $U\Omega=\Omega$ and
$\alpha (X)=UXU^{*}$ for all $X\in \alga$.  In particular,
$U\alg{B}U^{*}=\alg{B}$ and by continuity $U\alg{B}''U^{*}=\alg{B}''$.
For any $Z\in \alg{B}$, $U(Z\Omega )=UZU^{*}\Omega \in [\alg{B}\Omega
]$ and therefore $[U,P]=0$.  Thus, $U(ZP)U^{*}=(UZU^{*})P \in
\alg{B}''P$ for any $Z\in \alg{B}''$.  Thus,
$U\alg{B}^{\Omega}U^{*}=\alg{B}^{\Omega}$, and $\alpha
(\alg{B}^{\Omega}\cap \algae )=U(\alg{B}^{\Omega}\cap \algae
)U^{*}=\alg{B}^{\Omega}\cap \alg{A}$.  Therefore $\alg{B}^{\Omega}\cap
\alga$ is definable in terms of $\alg{R}$ and $\omega$.

(b.) We now show that $U_{t}\in \alg{B}^{\Omega}\cap \alg{A}$ for all
$t\in \mathbb{R}$.  Clearly $U_{t}\in \alg{A}$ since $U_{t}=\exp \{
-itH\}$ and $H$ is a finite linear combination of elements in $\{
U_{a}\otimes U_{b}:a,b\in \mathbb{R} \}$.  Now, for any $Z\in
\alg{B}$, we have shown that $U_{t}Z\Omega=ZU_{t}\Omega=Z\Omega$.  
Thus, $U_{t}$ acts
like the identity on the subspace $[\alg{B}\Omega ]$ of $\hil{H}$.
If $P$ is again used to denote 
the orthogonal projection onto $[\alg{B}\Omega ]$,
then $U_{t}P=P\in \alg{B}''P$; and therefore $U_{t}\in
\alg{B}^{\Omega}$.

(II.) We prove that there is a unique subalgebra of $\alg{A}$ that
maximally satisfies conditions 1.--4. of the theorem.  This
result turns on the following key fact: For \emph{any} representation
$(\pi ,\hil{H})$ of $\alg{A}$, the von Neumann algebra $\pi ( \{
U_{a}\otimes U_{b}:a,b\in \mathbb{R} \} )''$ is maximal abelian in
$\bh$ \cite[Thm.~I.6]{fannes}.  Thus, in our present notation
(suppressing the representation mapping), $\{ U_{a}\otimes
U_{b}:a,b\in \mathbb{R} \}''$ is maximal abelian.

Let $\alg{F}=\{ U_{a}\otimes U_{b}:a,b\in \mathbb{R}\}$, and let $P$
denote the orthogonal projection onto $[\alg{F}\Omega ]$.  Since
$\alg{F}$ leaves $[\alg{F}\Omega ]$ invariant, $P\in \alg{F}'$.  Let
$\alg{F}^{\Omega}$ denote the subalgebra of $\bh$ given by
\begin{equation}
\alg{F}^{\Omega}=(I-P)\bh (I-P)\oplus \alg{F}''P .
\end{equation}
It is clear that $\alg{F}\subseteq \alg{F}^{\Omega}\cap \alg{A}$.
Thus, $\alg{F}^{\Omega}\cap \alg{A}$ satisfies conditions 2. and 4. of
the theorem.  Since $\alg{F}''P$ is abelian, $\alg{F}^{\Omega}\cap
\alg{A}$ satisfies 1. \cite[Prop.~2.2]{beables}.  And, since $\alg{F}^{\Omega}\cap
\alg{A}$ is constructed out of elements invariant under automorphisms 
that preserve the EPR state and $\alg{Q}_{1}$, it satisfies 3. For 
maximality, we must 
show that $\alg{F}^{\Omega}\cap \alg{A}$ contains any other
subalgebra $\alg{B}\subseteq\alg{A}$ satisfying 1.--4.

By hypothesis, $\alg{B}$ is a subalgebra of $\alg{A}$ such that
$\omega |_{\alg{B}}$ is classical and $\alg{F}\subseteq \alg{B}$.
(Henceforth, we shall not actually need $\alg{B}$'s satisfaction of
condition 3.) Then, $\omega |_{\alg{B}''}$ is classical and
$\alg{F}''\subseteq \alg{B}''$.  Since $\alg{F}''$ is maximal abelian,
$\alg{F}''=\alg{F}'$.  Thus, $P\in \alg{F}'=\alg{F}''\subseteq
\alg{B}''$.  Let $A$ be an arbitrary element of $\alg{B}$.  Then,
$A\Omega =AP\Omega =PA\Omega$ since $A,P\in \alg{B}''$ and since
$\omega |_{\alg{B}''}$ is classical.  Thus, $A$ leaves $[\alg{F}\Omega
]$ invariant, and $A=(I-P)A(I-P)+AP$.  Furthermore, for any $R\in
\alg{F}''$, $[AP,RP]=0$.  Since $\alg{F}''P$ is a maximal abelian
subalgebra of $P\bh P$, it follows that $AP\in \alg{F}''P$.
Therefore, $A\in \alg{F}^{\Omega}\cap \alg{A}$.  Since $A$ was an
arbitrary element of $\alg{B}$, it follows that $\alg{B}\subseteq
\alg{F}^{\Omega}\cap \alg{A}$.  \end{proof}

\section{Conclusion}
We have shown that Bohr's reply to EPR is a logical consequence of
four requirements: (1.) {\it Empirical Adequacy:} When an observable
is measured, it possesses some value in accordance with the
probabilities determined by the quantum state. (2.) {\it Classical
  Description:} Properties $P$ and $P'$ can be simultaneously real in
a quantum state only if that state can be represented as a joint
classical probability distribution over $P$ and $P'$.  (3.) {\it
  Objectivity:} Elements of reality must be invariants of those
symmetries that preserve the defining features of the measurement
context.  (4.) {\it Maximality:} Our description should be maximal,
subject to the prior three constraints.  Obviously, these requirements
have nothing to do with the verifiability criterion of meaning or with
other central positivistic doctrines.  Thus, Bohr's reply to EPR does
not require a shift towards positivism.

Nonetheless, our reconstruction of Bohr's reply does not in itself
constitute an argument for the superiority of Bohr's point of view
over EPR's more ``realist'' point of view, which rejects the claim
that the reality of a system can be constituted ``from a distance.''
However, we wish to emphasize that Bohr is not so much concerned with
what is \emph{truly} real for the distant system as he is with the
question of what we would be \emph{warranted in asserting} about the
distant system from the standpoint of classical description.  In
particular, Bohr argues that in certain measurement contexts we are
warranted in attributing certain elements of reality to distant
(unmeasured) systems. He also claims, however, that if we attempt to
make \emph{context-independent} attributions of reality to these
distant systems, then we will come into conflict with the experimental
record.

Moreover, as Bohr himself might have claimed, a similar sort of
context-dependence already arises in special relativity.  In
particular, an inertial observer is warranted in saying that any two
events that are orthogonal to his worldline at some worldpoint are
simultaneous.  However, if we attempt to make
\emph{context-independent} attributions of simultaneity to distant
events --- where the ``context'' is now set by the observer's frame of
reference --- then we will run into conflicts with the experimental
record.

Of course, a proper defense of Bohr's point of view would require much
more space than we have here.  However, we have supplied ample
justification for the claim that Bohr's reply to EPR --- and his
philosophy of quantum theory in general --- deserves a more fair
treatment than it has recently received.

\bigskip {\it Acknowledgment:} We would like to thank Jeremy
Butterfield for a close reading of an earlier draft.


\begin{thebibliography}{99}
  
\bibitem{bohr35} Bohr, N. (1935) Can quantum-mechanical description of
  physical reality be considered complete? {\em Physical Review} {\bf
    48}, 696--702.
  
\bibitem{bohrletter} Bohr, N. (1935) Quantum mechanics and physical
  reality, {\em Nature} {\bf 136}, 65.
  
\bibitem{bohr49} Bohr, N. (1949) Discussion with {E}instein on
  epistemological problems in atomic physics, in P. Schilpp (ed.),
  {\em Albert Einstein: Philosopher-Scientist}, Tudor, NY, 201--241.
  
\bibitem{bub0} Bub, J. (1995) Complementarity and the orthodox
  (Dirac-von Neumann) interpretation of quantum mechanics, in R.
  Clifton (ed.), {\em Perspectives on Quantum Reality}, Kluwer, NY,
  211--226.
  
\bibitem{bub} Bub, J. (1997) {\em Interpreting the Quantum World},
  Cambridge University Press, NY.
  
\bibitem{bc} Bub, J. and Clifton, R. (1996) Uniqueness theorem for
  ``no-collapse'' interpretations of quantum mechanics, {\em Studies
    in the History and Philosophy of Modern Physics} {\bf 27},
  181--219.
  
\bibitem{fuchs} Caves, C., Fuchs, C. and Schack, R. (2001) Making good
  sense of quantum probabilities. E-print: quant-ph/0106133 .
  
\bibitem{kd} Clifton, R. (1995) Independently motivating the
  {K}ochen-{D}ieks modal interpretation of quantum mechanics, {\em
    British Journal for the Philosophy of Science} {\bf 46}, 33--57.
  
\bibitem{rob} Clifton, R. (1996) The properties of modal
  interpretations of quantum mechanics, {\em British Journal for the
    Philosophy of Science} {\bf 47}, 371--398.
  
\bibitem{rindler} Clifton, R. and Halvorson, H. (2001) Are Rindler
  quanta real?  Inequivalent particle concepts in quantum field
  theory, {\em British Journal for the Philosophy of Science} {\bf
    52}, 417--470.  E-print: quant-ph/0008030 .
  
\bibitem{epr} Einstein, A., Podolsky, B., and Rosen, N. (1935) Can
  quantum-mechanical description of physical reality be considered
  complete?  {\em Physical Review} {\bf 47}, 777--780.
  
\bibitem{fannes} Fannes, M., Verbeure, A., and Weder, R. (1974) On
  momentum states in quantum mechanics, {\em Ann. Inst.  Henri
    Poincar{\'e}} {\bf 20}, 291--296.
  
\bibitem{fine} Fine, A. and Beller, M. (1994) Bohr's response to EPR,
  in J.~Faye and H.~Folse (eds.), {\em Niels Bohr and Contemporary
    Philosophy}, Kluwer, NY, 1--31.
  
\bibitem{lmp} Halvorson, H. (2000) The Einstein-Podolsky-Rosen state
  maximally violates Bell's inequalities, {\em Letters in Mathematical
    Physics} {\bf 53}, 321--329.  E-print: quant-ph/0009007 .
  
\bibitem{beables} Halvorson, H. and Clifton, R. (1999) Maximal beable
  subalgebras of quantum mechanical observables, {\em International
    Journal of Theoretical Physics} {\bf 38}, 2441--2484.  E-print:
  quant-ph/9905042 .
  
\bibitem{how79} Howard, D. (1979) {\em Complementarity and Ontology:
    Niels Bohr and the problem of scientific realism in quantum
    physics},  PhD Dissertation, Boston University.
  
\bibitem{how94} Howard, D. (1994) What makes a classical concept
  classical?, in J.~Faye and H.~Folse (eds.), {\em Niels {B}ohr and
    {C}ontemporary {P}hilosophy}, Kluwer, NY, 201--229.
  
\bibitem{how00} Howard, D. (2000) A brief on behalf of {B}ohr,
  University of Notre Dame, manuscript.
  
\bibitem{kr} Kadison, R. and Ringrose, J. (1997) {\em Fundamentals of
    the Theory of Operator Algebras}, American Mathematical Society,
  Providence, RI.
  
\bibitem{malament} Malament, D. (1977) Causal theories of time and the
  conventionality of simultaneity, {\em No\^{u}s} {\bf 11}, 293--300.
  
\bibitem{pedersen} Pedersen, G. (1978) {\em {$C^{*}$}-Algebras and
    their Automorphism Groups}, Academic Press, NY.
  
\bibitem{petersen} N.~Bohr quoted in A.~Petersen, {\it Bulletin of the
    Atomic Scientists} {\bf 19}, 8--14.
  
\bibitem{redhead} Redhead, M. (1989) {\em Incompleteness, Nonlocality,
    and Realism}, Oxford University Press, NY.
  
\bibitem{ruark} Ruark, A. (1935) Is the quantum-mechanical description
  of physical reality complete?  {\em Physical Review} {\bf 48},
  466--467.
  
\bibitem{sakai} Sakai, S. (1971) {\em $C^{*}$-Algebras and
    $W^{*}$-Algebras}, Springer, NY.

\end{thebibliography}
\end{document}